\documentclass[twocolumn,aps,prl,floatfix,superscriptaddress,showkeys,preprintnumbers]{revtex4}

\usepackage{amsmath}    
\usepackage{graphicx}   	
\usepackage{verbatim}   	
\usepackage{color}      	
\usepackage{subfigure}  	
\usepackage{hyperref}   	
\usepackage{bm}

\begin{document}
\graphicspath{{./Figures/}}

\def\eps{\varepsilon}
\def\epsCr{\varepsilon_\mathrm{cr}}
\def\sigmaCr{\sigma_\mathrm{cr}}
\def\p{\partial}
\def\k{\bm{k}}
\def\etal{{\it~et~al.~}}

\title{Numerical simulations of aggregate breakup in bounded \\Ê and unbounded turbulent flows}

\author{Matthaus U. Babler}\email{babler@kth.se}
\affiliation{Department of Chemical Engineering and Technology, KTH Royal Institute of Technology, SE-10044 Stockholm, Sweden}

\author{Luca Biferale} 
\affiliation{Dept. of Physics and INFN, University of Rome Tor Vergata, Via della Ricerca Scientifica 1, 00133 Roma, Italy}

\author{Luca Brandt}
\affiliation{Linn{\'e} FLOW Centre and SeRC (Swedish e-Science Research Centre), KTH Mechanics, SE-10044 Stockholm. Sweden}

\author{Ulrike Feudel}
\author{Ksenia Guseva}
\affiliation{Theoretical Physics/Complex Systems, ICBM, Carl von Ossietzky University, Oldenburg, Oldenburg, Germany}

\author{Alessandra S. Lanotte}
\affiliation{ISAC-CNR and INFN, Sez. Lecce, 73100 Lecce, Italy}

\author{Cristian Marchioli}
\affiliation{Dept. Electrical, Management and Mechanical Engineering, University of Udine, 33100 Udine, Italy}
\affiliation{Dept. Fluid Mechanics, CISM, 33100 Udine, Italy}

\author{Francesco Picano}
\affiliation{Linn{\'e} FLOW Centre and SeRC (Swedish e-Science Research Centre), KTH Mechanics, SE-10044 Stockholm. Sweden}
\affiliation{Dept. Industrial Engineering, University of Padova, Via Venezia 1, 35131 Padova, Italy}

\author{Gaetano Sardina}
\affiliation{Linn{\'e} FLOW Centre and SeRC (Swedish e-Science Research Centre), KTH Mechanics, SE-10044 Stockholm. Sweden}

\author{Alfredo Soldati}
\affiliation{Dept. Electrical, Management and Mechanical Engineering, University of Udine, 33100 Udine, Italy}
\affiliation{Dept. Fluid Mechanics, CISM, 33100 Udine, Italy}

\author{Federico Toschi}
\affiliation{Dept. Applied Physics, Eindhoven University of Technology, 5600 MB Eindhoven, the Netherlands}
\affiliation{IAC, CNR, Via dei Taurini 19, 00185 Roma, Italy}

\date{\today}

\begin{abstract}
Breakup of small aggregates in fully developed turbulence is studied
by means of direct numerical simulations in a series of typical
bounded and unbounded flow configurations, such as a turbulent channel
flow, a developing boundary layer, and homogeneous isotropic
turbulence. The simplest criterion for breakup is adopted, 
whereby aggregate breakup occurs when the local hydrodynamic stress
$\sigma\sim\eps^{1/2}$, with $\eps$ being the energy dissipation at the
position of the aggregate, overcomes a given threshold $\sigmaCr$,
which is characteristic for a given type of aggregate. Results show that the
breakup rate decreases with increasing threshold. For small
thresholds, it develops a scaling behaviour among the different flows. For high thresholds, the
breakup rates show strong differences between the different flow
configurations, highlighting the importance of non-universal mean-flow
properties. To further assess the effects of flow inhomogeneity and
turbulent fluctuations, the results are compared with those obtained in a
smooth stochastic flow. Furthermore, we discuss the limitations and
applicability of a set of independent proxies.
\end{abstract}

\preprint{Version accepted for publication (postprint)}

\maketitle


%
%
\section{Introduction}
Particles in the colloidal and micrometre size range have a strong
tendency to stick together and form aggregates that, depending on the
type of particle and the environment, may undergo further
transformations such as coalescence or sintering to form compact
structures. Turbulence in the suspending fluid has a distinct
influence on the aggregation process: it leads to an enhancement of
the rate at which aggregates grow, e.g. by facilitating collisions between particles
\citep{Kusters1997,BablerLangmuir2010,BrunkKoch_jfm1998,Reade_Collins2000},
and it induces breakup of the formed aggregates
\citep{Flesch1999,Kobayashi1999,DerksenAICHE2012,Selomulya2002}. Breakup
is an important phenomenon in aggregation processes
\citep{YuanFarnood_2010,Bubakova2013,Tao_PowderTech2006}, as it is one
of the two main mechanisms that can interrupt aggregate growth in a
destabilized suspension of infinite extent (the other mechanism being
sedimentation, which removes large aggregates from the suspension).
This is experimentally evidenced by monitoring the evolution of the
aggregate size in a stirred suspension of destabilized particles
\citep{SoosJCIS2008,BiggsHounslow2003}. Starting from primary
particles, the aggregate size first undergoes a rapid increase before
levelling off to a plateau, where aggregation and breakup balance each
other. At this point, increasing the stirring speed increases the
magnitude of breakup which results in a rapid relaxation of the
aggregate size to a new plateau at a smaller size.

Breakup of aggregates has attracted considerable attention in the
literature
\citep{ZaccoPRE2009,OConchuir2013,Potanin1993,Eggersdorfer2010}. The
aggregate strength is experimentally measured by immersing
pre-prepared aggregates into a sufficiently diluted flow and measuring
the size and structure of the fragments that do not undergo further
breakup
\citep{SonntagRussel_1986a,Kobayashi1999,SoosNozzle2010,Harshe_Exp2011}.
Assuming that the largest fragments make up the remainder of the original
clusters, this technique allows one to interpret the measured fragment size as the
aggregate strength. From such experiments it was found that the
typical aggregate size that can withstand breakup decreases with the applied hydrodynamic stress according to a power law or, expressed the other way round,
the aggregate strength decreases with increasing size. However,
the strength {\it per se} gives no information on the rate of breakup,
i.e. how fast the number of aggregates decays in time. The rate of
breakup is a crucial quantity in the dynamics of aggregation
processes since it influences restructuring \citep{Selomulya2003} and
crucially controls the steady-state cluster size distribution
\citep{BablerJCIS2007}. Moreover, it is an important quantity for
modelling aggregation processes by means of population balance
equations, where breakup typically is described as a rate process
\citep{Soos_CES2006,MaerzFeudel2011,Flesch1999}.

Early models relating the breakup rate to the aggregate strength were
presented by \citet{Delichatsios1975} and \citet{Loginov1985} (for the
conceptually equivalent case of breakup of sub-Kolmogorov droplets): these were
followed by the exponential model of \citet{Kusters_thesis} and the
engulfment model of \citet{BablerJFM2008}. The basic principle is that
an aggregate suspended in a turbulent flow is subject to a fluctuating
hydrodynamic stress that only intermittently overcomes the critical
stress required to break the aggregate. The breakup rate is then
derived from the time it takes for an aggregate to experience such
stress.

Describing how long it takes for an aggregate to experience a breaking
stress is not an easy task, as the fluctuations in the stress
experienced by an aggregate are controlled not only by turbulent
fluctuations but also by the way the moving aggregate samples these
fluctuations. Accordingly, predicted breakup rates vary greatly among
the different models and even lead to contradictory results: in the
limit of very weak aggregates the exponential model of
\citet{Kusters_thesis} predicts a constant breakup rate, while the
engulfment model leads to a diverging breakup rate
\citep{BablerJFM2008}. In \citet{BablerPRE2012}, direct numerical
simulation (DNS) was used to obtain Lagrangian trajectories of
point-like aggregates released into homogeneous and isotropic
turbulence (HIT): trajectories were followed until the aggregates
experienced a stress that is able to break them. The breakup rate
measured in this study showed some characteristic properties that were
only partially captured by earlier models. In particular, for small
values of the aggregate strength, the breakup rate follows a power
law, whereas in the opposite limit of the aggregate strength becoming
large, the breakup rate decreases with a sharp superexponential cut
off. While the behaviour at large aggregate strength was well captured
by the engulfment model, the power-law behaviour was overestimated by
both the engulfment model and the exponential model.

More recently, a similar analysis \citep{DeBonaJFM2013} was performed by combining data
obtained from a DNS of HIT with Discrete Element Methods based on
Stokesian dynamics, modelling in detail the internal stresses while
the aggregates are moving in the turbulent field. This more detailed
analysis confirmed the power-law behaviour of the breakup rate in the
limit of small aggregate strength, while in the opposite limit of
large strength, a slightly slower dropoff was observed, due to the
role of internal stresses and aggregate orientation in the flow.\\

Most of the works discussed so far considered aggregate dynamics in
homogeneous and isotropic turbulence, which for real turbulent flows
holds only on a sufficiently small length scale and for distances far
enough from the walls. The question thus arises as to what extent
results from homogeneous flows apply to real flows, which are strongly
influenced by their boundary conditions. With the aim of answering this question, in the present work we
investigate breakup of aggregates in wall-bounded flows, namely a
developing boundary layer flow (BLF) and a channel flow
(CF). Aggregate breakup is studied by means of numerical experiments
using the same methodology as in our previous work
\citep{BablerPRE2012}. Specifically, the aggregates are assumed to be small
with respect to the viscous length scale of the flow, and their
inertia negligible. Also, their concentration is assumed to be low, such
that the properties of the flow are not altered due to the presence of
the aggregates. This situation typically applies to
  aggregates in liquid media, such as in the aggregation of
  diluted polymeric latexes \citep{SoosJCIS2008} or in the transport
  of suspended solids in estuaries \citep{Fugate_estuary2003}. It may
  not apply to aggregates that are heavier (or substantially lighter) than the
  fluid and/or finite-size aggregates that are significantly larger
  than the viscous length scale, in which cases inertia becomes
  important. Furthermore, breakup is assumed to occur whenever the
hydrodynamic stress, taken as the local energy dissipation at the
position of the aggregate, exceeds a predefined threshold representing
the aggregate strength. This rule represents the simplest
  breakup criterion for aggregate breakup using a single parameter
  (the aggregate strength) to determine the occurrence of breakup and
  ignoring any accumulation of stress inside the aggregate.  In
future work, this criterion could be refined by introducing degrees of freedom for the internal
dynamics, leading to stretching-relaxation effects
similar to those in the cases of droplet deformation/breakup and the polymer
coil/stretch transition in turbulent flows
\citep{balkovsky1999,biferale2014,MaffettoneMinale98}.

In non-homogeneous flows the breakup rate depends on the spatial location at which at which the aggregates are released. Aggregates released in a calm region would first move to a more intense region where it becomes more likely that they will experience a stress that can break them: thus these aggregates, on average, would survive for longer than aggregates that are released directly into the more intense region. This makes the breakup rate in the former case smaller than that of the latter. These subtleties make a complete characterization of breakup in non-homogenous flows cumbersome. Therefore, we restrict our analysis of breakup in non-homogenous flows to some specific situations, i.e. for the two bounded flows we consider only the cases 
where aggregates are released close to the wall and far away from it.
Despite the strong non-homogeneity and the presence of a mean shear in
wall-bounded flows, the measured breakup rate in each of these cases
shows some remarkable similarities to the breakup rate in homogeneous
turbulence. To corroborate and better understand this behaviour, we
additionally consider a synthetic turbulent flow (STF), obtained by
stochastically evolving the Fourier modes of a random velocity
field. Measuring the breakup rate in this flow leads to similar power-law 
behaviour suggesting that the latter is caused by weak turbulent
fluctuations, which are well represented by Gaussian statistics and therefore
only weakly influenced by the flow's boundary conditions. The breakup
rate of strong aggregates, on the other hand, is substantially larger
in wall-bounded flows, as compared to homogeneous turbulence where
only rare intermittent bursts can break strong aggregates.

%
%
\section{Numerical experiments}

%
%
\subsection{Aggregate breakup in turbulent flows}\label{sec:method}
As in \citet{BablerPRE2012}, we consider a situation where preformed
aggregates are released at a given location into a stationary flow
containing no other particles. The flow is assumed to be diluted
such that its statistical properties are not affected by the presence
of the aggregates (i.e. one-way coupling between the fluid and the
particulate phase). Furthermore, the aggregate density is assumed to
be close to the fluid density, and the aggregate size is assumed to be small relative to the dissipative length scale of the flow but large enough
for Brownian motion to be negligible. This is typical for
  polymeric colloids in liquid media. \citet{SoosJCIS2008} studied the
  aggregation of polystyrene particles in a diluted flow for which
  they found aggregate sizes in the range of 10 to 40 $\mu$m,
  depending on the stirring speed. The corresponding dissipative length scale is
  reported to vary on average between $30$ and $120$ $\mu$m, and the
  aggregate density is estimated as $1.02$ g/cm$^3$ (by assuming
  compact aggregates with a porosity of 40\% and taking the bulk
  density of polystyrene to be $1.05$ g/cm$^3$).  For such aggregates,
  the Stokes time $\tau_p=(2\rho_p+\rho)r^2/(9\rho\nu)$, where $\rho$
  is the fluid density, $\rho_p$ the aggregate density, $\nu$
  the kinematic viscosity and $r$ the aggregate radius, varies
  between $0.01$ and $0.15$ ms. Hence $\tau_p$ is small relative
  to the fastest turbulent time scales of the employed flows, reported
  to vary on average between $\tau_\eta=0.7$ and $10$ ms. The Stokes number
  defined as the ratio between the two time scales is of the order of
  $St\sim 10^{-2}$, which implies very small inertia.
  
Although the
  time scale and length scale of turbulent fluctuations are subject to variations and may assume
  substantially smaller values during intense turbulent events \citep{Biferale_pof2008}, which
  consequently would cause some inertial effects on the aggregate
  motion, here we consider the case where the aggregates have
  negligible inertia, i.e. we assume $St \sim 0$ throughout the
  flow. On the one hand, this allows us to treat the aggregates as if
  they were tracers: hence, despite their finite size, the aggregate
  trajectory is simply described by
\begin{equation}\label{eq:dxdt}
\frac{\mathrm{d}\bm{x}(t)}{\mathrm{d}t}=\bm{u}(\bm{x}(t),t)\,.
\end{equation}
where $\bm{x}(t)$ is the position of the centre of mass of
the aggregate at time $t$ and $\bm{u}(\bm{x},t)$ is the velocity
field. On the other hand, this assumption identifies the breakup mechanism to be
due only to hydrodynamic shear acting on the aggregate.

We define breakup as a singular event in time, i.e. there is an exact
moment in time when an aggregate turns from being intact into being
broken. We assume that this happens when the local stress acting on
the aggregate exceeds a critical stress $\sigma_\mathrm{cr}$
\citep{BablerPRE2012,BablerJFM2008}, i.e. we consider the
  limit of highly brittle aggregates which is believed to hold for
  small and compact aggregates made of materials that form stiff
  bonds, such as certain polymeric latexes \citep{ZaccoPRE2009}. In
  this limit, the time for accumulating the stress is small compared with
  the time over which the stress is applied, so that with respect
  to the time scale of the stress fluctuations breakup occurs {\it
    instantanously}. The critical stress is a characteristic of the
aggregate under consideration, i.e. $\sigma_\mathrm{cr}$ is a function of the
aggregate properties such as size, structure, type of constituent
particles, and chemical environment. Of these variables, the size
of the aggregate is the most crucial. A large body of experimental
\citep{SonntagRussel_1986a,SoosNozzle2010,Harshe_Exp2011}, numerical
\citep{Harshe2012,Eggersdorfer2010,BeckerBriesen2009,Vanni_Langmuir2011},
and theoretical studies \citep{ZaccoPRE2009} suggests a power-law
dependency of the form
\begin{equation}
\label{eq:sizerel}
\sigma_\mathrm{cr}\sim r^{-q}, \quad \mathrm{respectively} \quad
\sigma_\mathrm{cr}\sim \xi^{-q/d_f},
\end{equation}
where $\xi\sim r^{d_f}$ is the number of primary particles
constituting the aggregate, $d_f$ is the aggregate fractal dimension,
and $q$ is a scaling exponent that depends on the aggregate
structure. For dense but non-compact aggregates, \citet{ZaccoPRE2009}
give $q=[9.2(3-d_f)+1]/2$ in good agreement with experiments
\citep{Harshe_Exp2011,SoosAICHE2013}.

\begin{figure}\begin{center}
\includegraphics[width=8cm]{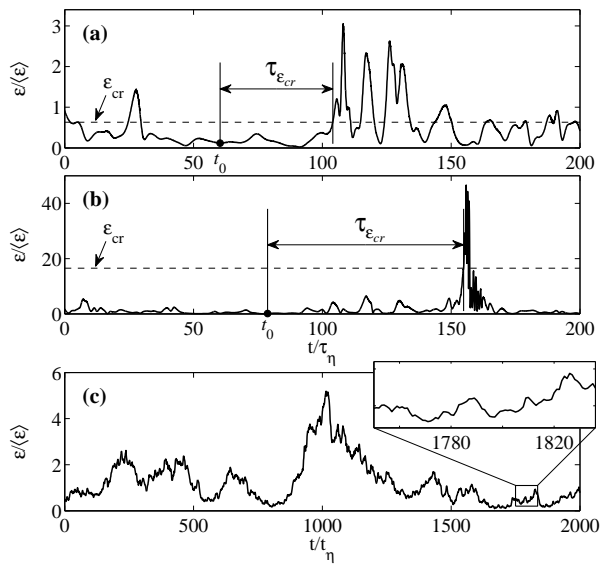}
\caption{\label{fig:F100} Definition of the exit-time, $\tau_{\epsCr}$
  (see text) for two typical trajectories in a homogeneous and
  isotropic flow: ({\it a}) Time series of energy dissipation along an
  aggregate trajectory for a low-turbulent intensity trajectory; ({\it
    b}) the same but in the presence of a strongly turbulent
  burst. The dashed line indicates the critical dissipation.  ({\it
    c}) A typical evolution of the energy dissipation for an aggregate
  evolving in a synthetic turbulent flow. Notice the absence of
  strong fluctuations in the latter case. The horizontal axis is
  normalized by the Kolmogorov time scale
  $\tau_\eta=(\nu/\langle\eps\rangle)^{1/2}$.}
\end{center}\end{figure}

The hydrodynamic stress acting on an aggregate is
$\sigma\sim\mu(\eps/\nu)^{1/2}$, where $\mu$ is the dynamic viscosity
and $\eps$ is the local energy dissipation rate, defined as
\begin{equation}\label{eq:edr}
\eps=2\nu s_{ij}s_{ij}\,,
\end{equation}
with $s_{ij}=\frac{1}{2}(\p u_i/\p x_j + \p u_j/\p x_i)$. Thus,
strong fluctuations of $\eps$ control the fluctuations of the stress
and therefore the occurrence of breakup events. This translates into a
picture where an aggregate upon release moves through the flow until
the local dissipation exceeds a threshold value $\epsCr \sim
[\sigma_\mathrm{cr}(\xi)]^2$ causing it to break up. Hence it is
crucial to control the typical time for which the aggregate experiences a local
stress below the critical value, what we call the exit-time. In
figure \ref{fig:F100}({\it a}) we show schematically the way in which we
propose to estimate the breakup rate, using a real example taken from
the evolution of one aggregate. In the figure, we show the time series
of kinetic energy dissipation along an aggregate trajectory and the
procedure followed to define the exit time. An aggregate
released at a time $t_0$ moves with the flow for a time
$\tau_{\epsCr}$ after which the local dissipation exceeds for the
first time the critical threshold $\epsCr$ (indicated by the dashed
line in figure \ref{fig:F100}({\it a})) where the aggregate breaks
up. The first crossing of $\epsCr$ thus defines the exit time,
$\tau_{\epsCr}$, which is the basic quantity for determining breakup
rates.

To measure the exit time of aggregates, the following protocol is
applied \citep{BablerPRE2012}. (i) At a time $t_0$, a given number of
aggregates is released at a random location within a domain $\Omega$
of a stationary flow. (ii) Aggregates released at a point where the
local dissipation exceeds $\epsCr$ are ignored, as breakup would have
already occurred before the aggregates could reach that
point. (iii) Each of the remaining aggregates is followed over time
until the local dissipation exceeds the critical dissipation
$\epsCr$: the time lag from release tol the breakup defines the
exit time $\tau_{\epsCr}$ for that aggregate. (iv) Fragments formed upon
breakup of an aggregate are
  discarded. The breakup rate for the given threshold and domain of
release is then given by the inverse of the mean of the exit time,
computed as  the ensemble average over many time histories:
\begin{equation}\label{eq:def1}
f_{\epsCr}=\frac{1}{\langle\tau_{\epsCr}\rangle}\,.
\end{equation}

Equation (\ref{eq:def1}) provides a valid definition of the breakup
rate that is applicable to both homogeneous and non-homogeneous
flows. However, it is important to notice that its implementation
requires one to observe the particles for a sufficiently long time in
order to confidently estimate the mean exit time. This can be very
challenging for measurements made in the field or in a laboratory, and
for large values of $\epsCr$ that occur only rarely. Hence,
approximations to the breakup rate given by (\ref{eq:def1}) are
desirable. One such approximation, applicable to homogeneous flows,
is obtained by considering the diving time, defined as the time
lag in between two consecutive crossings of the critical dissipation
\citep{Loginov1985}. In homogeneous flows, the diving time can be
obtained using the Rice theorem for the mean number of crossings per
unit time of a differentiable stochastic process, leading to the
following proxy for the breakup rate \citep{BablerPRE2012}:
\begin{equation}\label{eq:Loginov}
f_{\epsCr}^{(E)}=
\frac{\int_0^\infty \mathrm{d}\dot\eps \ \dot\eps p_2(\epsCr,\dot\eps)}{\int_0^{\epsCr} \mathrm{d}\eps \ p(\eps)}\,,
\end{equation}
where $p_2(\eps,\dot\eps)$ is the joint probability density function
(p.d.f..) of the dissipation and its time derivative, $p(\eps)$ is the p.d.f.
of $\eps$ and the superscript '($E$)' stands for 'Eulerian', indicating
that the fragmentation rate is estimated without the need of Lagrangian
properties.

Another important and potentially useful approximation can be derived
by considering the time evolution of the number of aggregates. In the
case where breakup is driven by an uncorrelated force field, the
breakup rate can be written as
\begin{equation}\label{eq:NdN}
f_{\epsCr}^{(N)} = - \frac{\mathrm{d}\ln N_{\epsCr}(t)}{\mathrm{d}t}\,,
\end{equation}
where $N_{\epsCr}(t)$ is the number of aggregates at time $t$ after
their release. The latter is simply related to the exit-time
measurements described above by the relation
\begin{equation}\label{eq:NdN_exit}
N_{\epsCr}(t)/N_{\epsCr}(0)=1-\int_0^t \mathrm{d}\tau \ p_{\epsCr}(\tau)\,,
\end{equation}
where $N_{\epsCr}(0)$ is the number of aggregates successfully
released into the flow and $p_{\epsCr}(\tau)$ is the p.d.f. of the
exit time for a threshold $\epsCr$.

%
%
\subsection{Flow fields}

\subsubsection{Boundary layer flow}
\begin{figure}\begin{center}
\includegraphics[width=7cm]{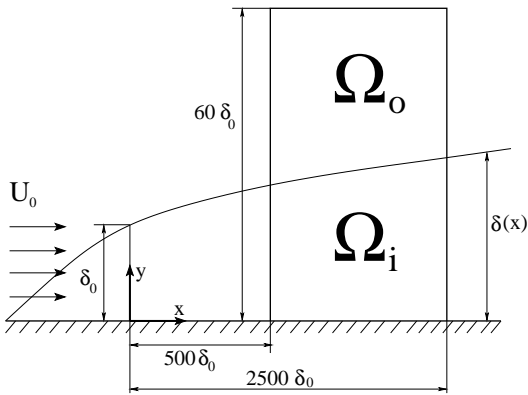}
\caption{\label{fig:bl} Schematic of the boundary layer flow. The two
  different seeding regions are labeled as $\Omega_i$ for the
  aggregates released inside the boundary layer, and $\Omega_o$ for
  aggregates released outside the boundary layer region. $\delta(x)$
  represents the geometric boundary layer thickness where the mean
  velocity is $99\%$ of the free stream velocity $U_0$; $\delta_0$
  denotes the boundary layer thickness in the inlet section of the
  computational domain. $x$ and $y$ denote the streamwise and
  wall-normal coordinate, respectively.}
\end{center}\end{figure}

We consider a zero-pressure-gradient flow, i.e. the case of a thin
flat plate immersed in a uniform steady stream of viscous fluid with
undisturbed characteristic velocity $U_0$. The no-slip boundary condition
is applied on the flat plate. The viscous stresses generated by the
flat plate retard the fluid elements close to the wall, so that the
fluid zone close to the flat plate has a velocity lower than the free
stream value $U_0$. The resulting flow is known as 'boundary layer flow'
(BLF). A sketch of the flow configuration is displayed in figure
\ref{fig:bl}. A typical measure of the boundary layer thickness is the
so-called geometric thickness, $\delta$, defined as the distance
perpendicular to the wall where the flow reaches $99\%$ of the
undisturbed free stream velocity.  It is known from experiments and
from simple dimensional arguments that the geometric thickness
increases as one moves downstream along the flat plate, implying that
the BLF is a spatially evolving flow with a strong inhomogeneity in
the wall-normal direction and a weaker evolution in the wall-parallel
directions. Different, though somehow equivalent measures of the
characteristic boundary layer thickness exist, such as the displacement
thickness $\delta^*$ and the momentum thickness $\theta$, which take into
account the mass and the momentum loss inside the boundary layer
\citep{schli}.

A DNS of the BLF was performed using the pseudospectral Navier-Stokes
solver SIMSON \citep{simson}. The computational domain has a size of
$(3000 \delta_0) \times (100 \delta_0) \times (120\delta_0)$ in
the streamwise, wall-normal and spanwise directions, where $\delta_0$
denotes the geometric boundary layer thickness at the inlet section
of the computational domain. The numerical resolution is $4096 \times
384$ Fourier modes in the wall-parallel plane and $301$ Chebishev
modes in the wall-normal direction. A localized forcing close to the inlet, random in
time and in the spanwise direction, is used to induce the
laminar-turbulent transition. The characteristic Reynolds number of
the flow, based on the momentum thickness $\theta$, ranges from
$Re_\theta=200$ at the inlet to $Re_\theta=2500$ at the end of the
domain. The resulting turbulent flow is analogous to that described
in \citet{SardinaJFM2012,SardinaFTAC}, where the transport and
dispersion of inertial particles in boundary layers is studied.

Aggregates are released in two regions: inside the boundary layer
(labelled $\Omega_i$ in figure~\ref{fig:bl}) and outside the boundary
layer (labelled $\Omega_o$ in figure~\ref{fig:bl}). The release regions
span the streamwise interval from $500\delta_0$ to $2500\delta_0$ so
as to avoid interferences due to the tripping forcing that promotes
transition to turbulence. The height of the total release region is
$60 \delta_0$, and the difference between $\Omega_i$ and $\Omega_o$ is
determined by the local geometric thickness of the boundary
layer. The latter ranges from $15 \delta_0$ at the beginning of the
release region to $45\delta_0$ at the end of the release
region. The release regions were chosen with regard to the
  spatial distribution of the mean energy dissipation, which, as shown
  below, shows strong variation in the wall-normal direction while
  exhibiting only a slow decay in the streamwise direction.  A total
of $2\times 10^6$ tracer aggregates are released into the
flow. Aggregate trajectories are obtained by integrating the velocity
field (equation (\ref{eq:dxdt})).
The fluid velocity and its spatial derivatives at the position of the
aggregate are quantified by means of a fourth-order spatial
interpolation, while a second-order Adams-Bashforth scheme is used for
integration of (\ref{eq:dxdt}). Further details about the numerics of
the Lagrangian tracking solver can be found in
\citet{SardinaJFM2012,SardinaJFM2012_2}.

An additional point concerns the characteristic energy
  dissipation used for normalizing the measured breakup rates. As the
  BLF is evolving in both the streamwise and wall-normal directions, to
  define a characteristic dissipation some additional constraints are
  required, i.e. a specific downstream distance or a spatial domain at
  which the characteristic dissipation is extracted. Here, we
  consider the inner release region $\Omega_i$ and take the
  characteristic energy dissipation, denoted by $\eps_0$, as the
  volume average over this domain: $\eps_0$ defined in this way is
  used for datasets of aggregates released both inside and outside the
  boundary layer. A summary of the properties of this flow is given in
  table \ref{tab:characteristics}.

%
%
\begin{table}
\begin{center}
\begin{tabular}{ll lll}
Flow &					& Release region				& $\eps_0$				& $\tau_0$ \\[3pt]
BLF	& $Re_\theta=2500$ 	& $\Omega_i$, $\Omega_o$	& $\langle\eps|\bm{x}\in\Omega_i\rangle$ & $(\nu/\eps_0)^{1/2}$ \\
CF	& $Re_\tau=150$		& $\Omega_c$, $\Omega_w$	& $\langle\eps\rangle$	& $(\nu/\eps_0)^{1/2}$	 \\
HIT	& $Re_\lambda\simeq 400$	& whole domain		& $\langle\eps\rangle$ 	& $(\nu/\eps_0)^{1/2}$ \\
STF	& $Re_\sigma=300$ 	& whole domain 			& $\langle\eps\rangle$ 	& $t_\eta$
\end{tabular}
\caption{\label{tab:characteristics} Parameters of the numerical
  experiments. $\eps_0$ and $\tau_0$ are the characteristic energy
  dissipation and the timescale used to normalize the data. In the
  boundary layer flow (BLF) and channel flow (CF), aggregates are
  released in two regions. In the BLF, aggregates are released inside
  the boundary layer $\Omega_i=\{500<x/\delta_0<2500, y < \delta(x)
  \}$, and outside the boundary layer $\Omega_o=\{500<x/\delta_0<2500,
  \delta(x) < y < 60\delta_0 \},$ where $\delta(x)$ and $\delta_0$ are
  the boundary layer thickness and the boundary layer thickness at the
  entrance to the computational domain, respectively. In the CF,
  aggregates are released in the center-plane $\Omega_c=\{y/h=0\}$ and
  in the wall region $\Omega_w=\{0.933 < |y/h| < 1 \}$ where $y$ is
  the wall normal coordinate and $h$ is the half channel height. In
  the isotropic and homogeneous turbulence (HIT) and in the synthetic
  flow (STF), aggregates are released homogeneously. }
\end{center}
\end{table}

\subsubsection{Channel flow}
\begin{figure}\begin{center}
\includegraphics[width=7cm]{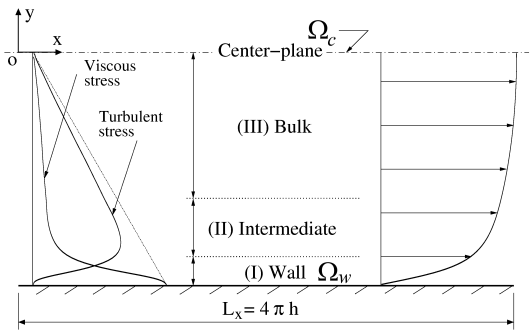}
\caption{\label{fig:sketch_CF} Schematic of the channel flow. The two
  different seeding regions are labeled as $\Omega_c$ for aggregates
  released in the center-plane and $\Omega_w$ for aggregates released
  near the wall. On the left and right sides are shown the mean
  profiles of turbulent and viscous stresses, and the mean velocity
  profile, respectively. $h$ denotes the half channel height.}
\end{center}\end{figure}

The flow domain consists of two infinite flat parallel plates, a
distance $2h$ apart. The origin of the coordinate system is located at
the centre of the channel, and the $x$, $y$ and $z$ axes represent
the streamwise, wall-normal and spanwise directions,
respectively. Periodic boundary conditions are imposed on the fluid
velocity field in homogeneous directions ($x$ and $z$) while no-slip
boundary conditions are imposed at the walls. The size of the
computational domain is $L_x \times L_z \times L_y = 4 \pi h \times 2
\pi h \times 2h$.  The flow is non-reactive, isothermal and
incompressible (low Mach number). The shear Reynolds number is
$Re_{\tau}=u_{\tau} h / \nu = 150$ \citep{MarchioliIJMF2008}, where
$u_{\tau}=\sqrt{\tau_w / \rho}$ is the shear velocity based on the
mean wall shear stress. The flow solver is based on the
Fourier-Galerkin method in the streamwise and spanwise directions, and
on a Chebishev-collocation method in the wall-normal direction. This
solver provides the spatial derivatives required to calculate fluid
dissipation along the aggregate trajectory according to (\ref{eq:edr})
with spectral accuracy.  A Lagrangian tracking code coupled with the
flow solver is used to calculate the path of each aggregate in the
flow. The aggregate equation of motion (\ref{eq:dxdt}) is solved using
a fourth-order Runge-Kutta scheme for time integration. Fluid velocity
and velocity derivatives at aggregate position are obtained using
sixth-order Lagrangian polynomials; at the wall, the interpolation
scheme switches to one-sided.  Further details on the numerical
methodology can be found in \citet{MarchioliIJMF2008} and
\citet{SoldatiIJMF2009}. A schematic of the flow is shown in figure
\ref{fig:sketch_CF}.

Following \citet{Pitton_pof2012} the flow domain is phenomenologically
divided into three regions: the wall region, the intermediate region and the
bulk region (see figure \ref{fig:sketch_CF}). The wall region
comprises a fluid slab with a thickness of 10 wall units. In this
region, the viscous stress (representing the mean fluid shear) is
maximal while the turbulent stress is close to zero. The intermediate
region extends up to $50$ wall units from the wall and is
characterized by the peak of the fluid Reynolds stresses. The bulk
region covers the central part of the channel where all wall stress
contributions drop to zero and turbulence is closer to homogeneous and
isotropic.  Breakup experiments are performed by releasing aggregates
in the wall region and at the centre-plane of the bulk region. The two
release regions are labelled $\Omega_w$ and $\Omega_c$ in figure
\ref{fig:sketch_CF}.  Within each of these release regions, $10^5$
aggregates are released and their trajectories are tracked and breakup
events detected. The characteristic energy dissipation $\eps_0$
used to normalize the breakup rate is taken as the volume average over
the whole flow domain: see table \ref{tab:characteristics}.

\subsubsection{Homogeneous turbulence (HIT)}
A DNS of three-dimensional incompressible Navier-Stokes turbulence was
performed on a triply periodic cubic box, with large-scale
statistically homogeneous and isotropic forcing. The external forcing
injects energy into the first low-wavenumber shells, by keeping their
spectral content constant \citep{Chen_1993}. The kinematic viscosity
is such that the Kolmogorov length scale is comparable to the grid
spacing; this choice ensures a good resolution of the small-scale
velocity dynamics. The Navier-Stokes equations are solved on a regular
grid, $2\pi$-periodic, by means of standard pseudospectral methods,
with time stepping done using a second-order Adams-Bashforth
algorithm. The grid has $2048^3$ points, and the Taylor scale-based
Reynolds number is $Re_{\lambda} \simeq 400$.  Lagrangian particle
velocities are obtained by a trilinear interpolation. Details on the
numerical integration can be found in \citet{Bec_JFM2010}. The
database for this study counts approximately $2\times 10^5$ tracer
trajectories. The characteristic dissipation for normalizing the
breakup rate, $\eps_0$, is taken as the mean dissipation over the whole
volume: see table \ref{tab:characteristics}.

\subsubsection{Synthetic Flow (STF)}
In order to better assess the importance of strong intermittent bursts
in the statistics of the energy dissipation felt by the aggregates, it
seems useful to study also the dynamical evolution in a STF, whose statistics can be controlled
a priori.  This flow is constructed to mimic properties of stationary HIT, but with an important and
crucial difference: it has Gaussian statistics for the
  velocity gradients. The STF is realized in a three dimensional
periodic box of size $L=2\pi$, with the velocity field written as a
Fourier series
\begin{equation}\label{eq:sm1}
\bm{u}(\bm{x},t)=\sum_{\bm{k}} \bm{\hat u^\prime_k}(t) e^{i\bm{kx}}\,.
\end{equation}
The Fourier coefficients satisfy $\bm{\hat u^\prime_{-k}}=\bm{\hat
  u_k^{\prime *}}$, where the asterisk indicates the complex conjugate. The
summation in (\ref{eq:sm1}) goes over $K=1, \ldots K_\mathrm{max}$
shells, each containing $N_K$ uniformly distributed wave vectors of
length $|\bm{k}|=K$. Incompressibility of $\bm{u}(\bm{x},t)$ is
ensured by taking $\bm{\hat u^\prime_k}$ as the projection of a
different vector $\bm{\hat u_k}$ on a plane perpendicular to
$\bm{k}$. The vector $\bm{\hat u_k}$ is evolved by a second-order
stochastic process originally proposed by \citet{Sawford1991} to
model Lagrangian dispersion.  Evolving $\bm{\hat u_k}$ by a second
order stochastic process results in a velocity field that is
differentiable in time, which is a crucial property for measuring
temporal statistics such as the exit time.  In the second-order
process, the spectral acceleration $\bm{\hat a_k}$ is given by the
following stochastic differential equation:
\begin{equation}
\mathrm{d}\bm{\hat a_k} = - \frac{\bm{\hat a_k}}{t_\eta} \mathrm{d}t - \frac{\bm{\hat u_k}}{t_\eta t_L}\mathrm{d}t + \sqrt{\frac{2\sigma^2_{\k}}{t_\eta^2 t_L}} \mathrm{d}\bm{W}\,, \label{eq:sm_da}
\end{equation}
where $\mathrm{d}\bm{W}$ is an incremental Wiener process, $t_\eta$
and $t_L$ are the time scales of acceleration and velocity,
respectively, and $\sigma_{\bm{k}}^2$ is the variance of a component
of $\bm{\hat u_k}$. Due to the isotropy of the flow field,
$\sigma_{\bm{k}}^2$ depends only on the modulus of $\bm{k}$,
i.e. $\sigma_{\bm{k}}^2=\sigma_K^2$, such that the energy carried by
all wave vectors of modulus $K$ is $E_K=\frac{3}{2}N_K\sigma_K^2$ and
the total energy is $E=\frac{3}{2}\langle
u^2\rangle=\sum_{K=1}^{K_\mathrm{max}}E_K$. The spectral velocity
$\bm{\hat u_k}$ is simply
\begin{equation}
\mathrm{d}\bm{\hat u_k} = \bm{\hat a_k}\mathrm{d}t\,. \label{eq:sm_du}
\end{equation}

In the present simulations, we set $K_\mathrm{max}=1$
\citep{BecJFM2005,Zhanow_Physica2011} and take the mean velocity
$\langle u^2\rangle^{1/2}$ to be small with respect to $L/t_L$. For this
choice of parameters, the Lagrangian properties are fully determined
by the evolution of the spectral coefficients. Following
\citet{Sawford1991}, the spectral acceleration decorrelates with $\sim
t_\eta$ while the integral scale of the spectral velocity is equal to
$t_L$.  This allows us to interpret $t_\eta$ as the equivalent of the
dissipative time scale in turbulence and, furthermore, motivates us to
estimate a small-scale Reynolds number for the STF, denoted by
$Re_\sigma$, as \citep{Sawford1991}
\begin{equation}
Re_\sigma \sim {t_L}/{t_\eta}\,,
\label{eq:Resynthetic}
\end{equation}
In this work, we set $Re_\sigma=300$ and use $t_\eta$ as the
characteristic time scale for normalizing the breakup rate; the
characteristic dissipation $\eps_0$ is taken to be the mean dissipation
(table \ref{tab:characteristics}). As in the other flows, the
aggregate trajectory is obtained by integrating the velocity field,
(\ref{eq:sm1}), while the local dissipation is obtained from
(\ref{eq:edr}), upon setting the value of the viscosity equal to unity.
For measuring breakup rates, several very long trajectories were
simulated, from which we then measured diving times. From the
diving time, the mean exit time was obtained from an exact relation
derived in \citet{BablerPRE2012}.  The breakup rate determined in this
way corresponds to the case where aggregates are released homogeneously
in the whole domain. The statistical database is as large as
$1.5\times 10^{6}$ diving events.

Before concluding this section, it is worth stressing the main
differences between the STF presented here and a realistic turbulent
flow. First, even though we can identify two different time scales in
the STF, the dissipation along an aggregate trajectory in STF will not
possess any anomalous and intermittent scaling: see
e.g. \citet{Biferale_trapping2005}. Second, the spatial configuration
of the STF is smooth and does not exhibit a Kolmogorov-like $-5/3$
spectrum. The former is particularly relevant and will be discussed
later in connection with the small efficiency of the STF to break strong
aggregates.

%
%
\section{Results}

%
%
\subsection{Properties of energy dissipation}
Energy dissipation plays a decisive role in the breakup of small
aggregates. Therefore, in this section we first explore the Lagrangian
and Eulerian properties of energy dissipation in the flows under consideration.

\begin{figure}\begin{center}
\includegraphics[width=8cm]{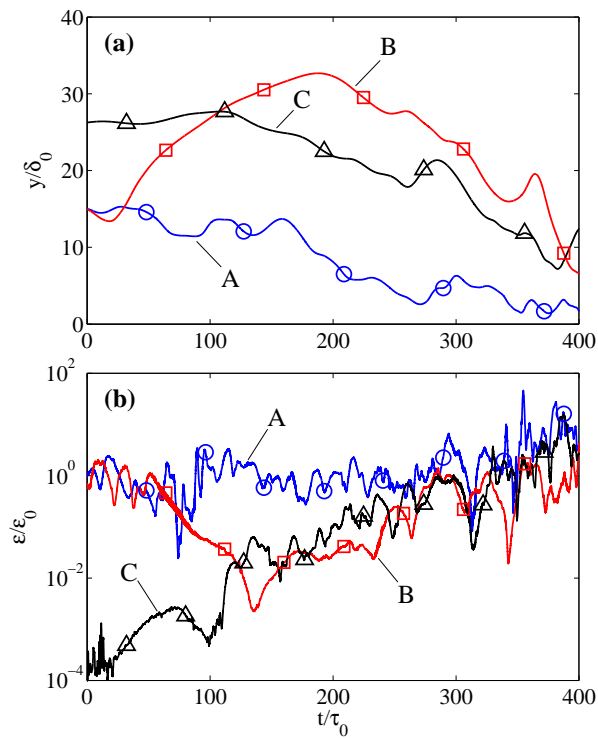}
\caption{\label{fig:F201} Time series of ({\it a}) wall normal
  distance and ({\it b}) energy dissipation along typical aggregate
  trajectories in the boundary layer flow. Trajectories A and B refer
  to aggregates released in $\Omega_i$, while trajectory C refers to
  an aggregate released in $\Omega_o$ (see figure \ref{fig:bl}).  Axis
  are normalized by $\eps_0$ and $\tau_0$ given in table
  \ref{tab:characteristics}.}
\end{center}\end{figure}

Figure~\ref{fig:F201} shows typical trajectories of tracer-like
aggregates in the BLF. Panel ({\it a}) shows time series of the wall-normal distance,
 while panel ({\it b}) shows the corresponding local
dissipation. For the cases shown in these plots, we assumed
  that the aggregates are infinitely strong such that they follow the
  trajectories without breakup (notice that the simple breakup
  criterion adopted in this work does not allow for determining the
  size and trajectories of the fragments formed). Among the three
trajectories shown, A and B are cases of aggregates released inside
the boundary layer, while C is a case of an aggregate released outside
the boundary layer. Within the observed time lag, aggregate A is
subject to strong fluctuations in dissipation that increase as it
moves downstream and as the aggregate comes closer to the wall. On the
other hand, aggregate B is first repelled from the boundary layer and
moves away from the wall: accordingly, the dissipation decreases and
fluctuations are rarer. Later, the aggregate is re-entrained into the
boundary layer, which causes dissipation to increase both in magnitude
and in the amplitude of fluctuations. The trajectory of aggregate B at this
later stage is thus similar to that of aggregate C, which is entrained into the
boundary layer after moving downstream for a certain distance.

From these apparently ad-hoc examples, it becomes clear that in the
presence of a mean flow, breakup events will be controlled by an
interplay between the mean flow properties and the relative
fluctuations around it.  For some aggregate histories the mean profile
will control the breakup process, whereas for others, breakup is
controlled by intense fluctuations of the local energy dissipation
around its mean value. As seen below, the balance between the two
depends strongly on the geometry of the flow configuration, on the
intensity of the turbulent fluctuations and also on where the
aggregates are released.

\begin{figure}\begin{center}
\includegraphics[width=8cm]{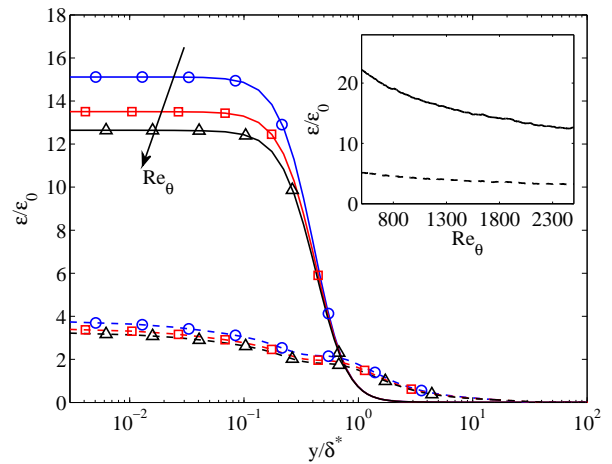}
\caption{\label{fig:F400} Time averaged energy dissipation in the
  boundary layer flow as a function of the wall normal distance at
  three downstream positions, characterized by $Re_\theta=1700$,
  $2100$, and $2500$, respectively. Solid lines: dissipation due to
  the mean flow; dashed lines: dissipation due to velocity
  fluctuations. The horizontal axis is normalized by the displacement
  thickness $\delta^*$, while the vertical axis is normalized by
  $\eps_0$ given in table \ref{tab:characteristics}.  Inset: Time
  averaged energy dissipation at the wall as a function of $Re_\theta$
  representing the streamwise direction. Solid and dashed lines have
  the same meaning as the main axes.}
\end{center}\end{figure}

The above discussion can be quantified by looking at the time-averaged
profiles of the energy dissipation in the BLF measured at three
downstream distances as shown in figure~\ref{fig:F400}, for the mean
flow and the fluctuating components. Close to the wall, dissipation
assumes large values that are dominated by the mean flow, as shown by
the solid curves in figure \ref{fig:F400}. Dissipation due to
turbulent velocity fluctuations (dashed curves) exhibits a flatter
profile that expands well beyond the boundary layer thickness
$\delta^*$. The decrease of dissipation in the streamwise direction is
small, as shown in the inset of figure \ref{fig:F400}, where we plot
the time-averaged dissipation at the wall as a function of
$Re_\theta$. This explains the relatively constant high magnitude of
energy dissipation seen by an aggregate moving within the boundary
layer (i.e. trajectory A in figure \ref{fig:F201}).

\begin{figure}\begin{center}
\includegraphics[width=8cm]{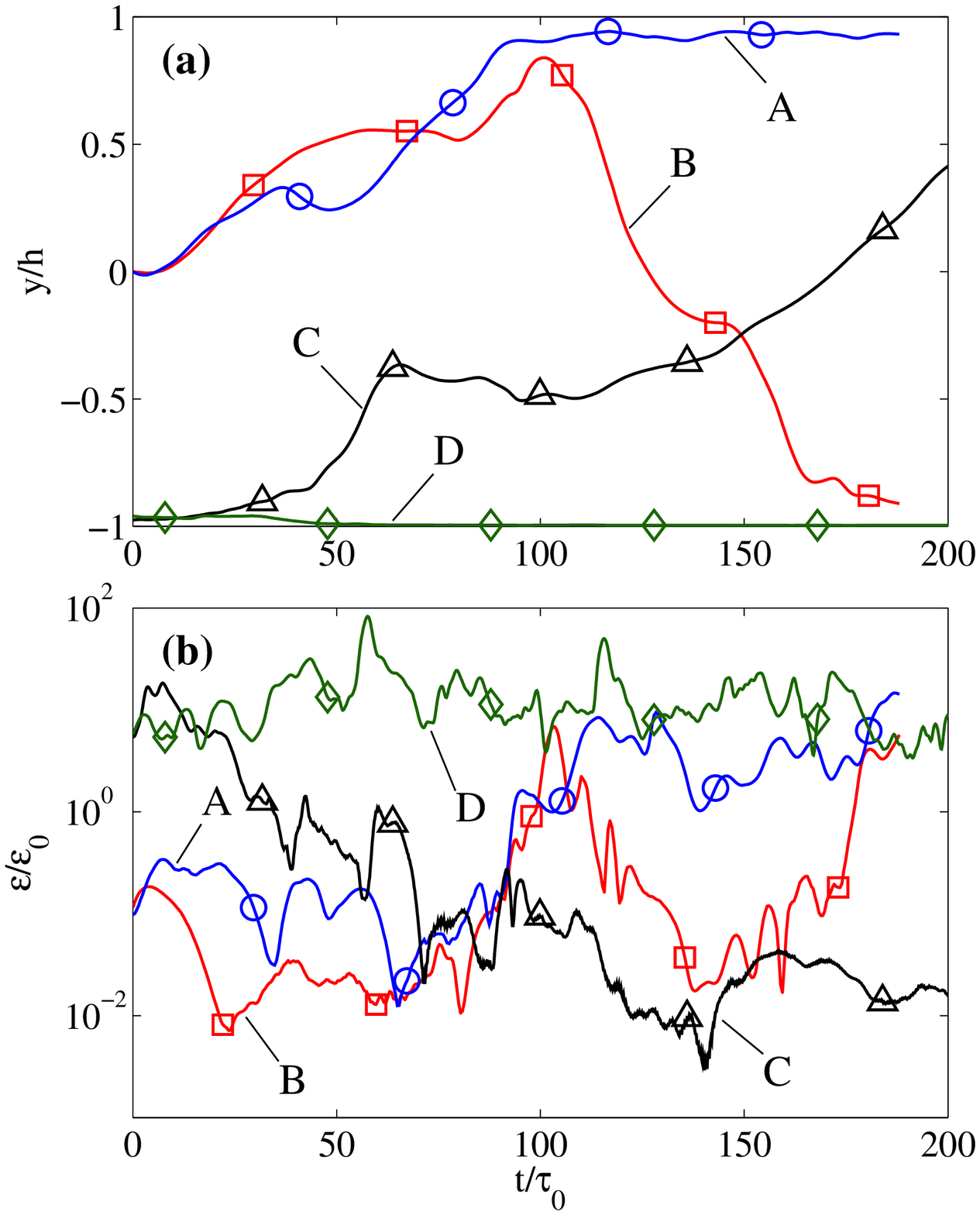}
\caption{\label{fig:F202} Time series of ({\it a}) wall normal
  distance and ({\it b}) energy dissipation along typical aggregate
  trajectories in the channel flow. Trajectories A and B refer to
  aggregates released in $\Omega_c$, while trajectories C and D refer to
  aggregates released in $\Omega_w$ (see figure
  \ref{fig:sketch_CF}). Axis are normalized by $\eps_0$ and $\tau_0$
  given in table \ref{tab:characteristics}.}
\end{center}\end{figure}

\begin{figure}\begin{center}
\includegraphics[width=8cm]{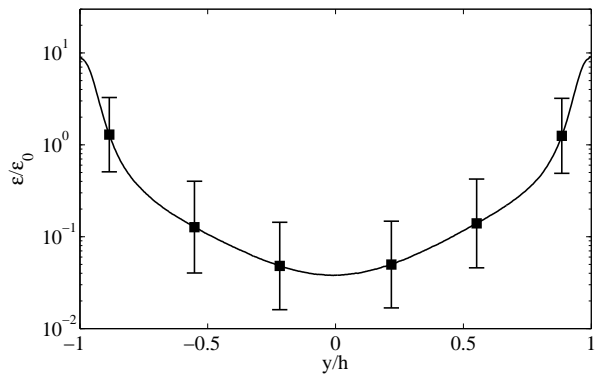}
\caption{\label{fig:F300} Mean energy dissipation conditioned on the
  wall normal distance in the channel flow. Error bars indicate the
  root-mean-square of the conditioned dissipation. The vertical axis
  is normalized by $\eps_0$ given in table \ref{tab:characteristics}.}
\end{center}\end{figure}

Figures~\ref{fig:F202} and \ref{fig:F300} show analogue data for the
CF. Here, aggregates A and B are released in the centre-plane of the
channel, while aggregates C and D are released in the wall region. The
aggregates released in the centre-plane gradually get entrained by
turbulent eddies which transport them to the walls. The entrainment
and transport to the wall cause an increase in the magnitude of
dissipation seen by the aggregate, while the fluctuations remain
persistent. Once they reach the wall, the aggregates have the tendency
to stay there for a relatively long time before being re-ejected into
the bulk flow. This is seen also for aggregates released close to the
wall: aggregate D stays close to the wall while aggregate C is
ejected into the bulk flow. From figure \ref{fig:F300},
where we plot the mean dissipation conditioned on the wall-normal
distance, it appears that aggregates are subject to high fluctuations of energy
dissipation even when staying in the bulk flow (i.e. away from the
walls). Similar to the BLF, dissipation assumes high values close to
the walls while fluctuations in dissipation, indicated by the error bars,
are intense throughout the channel.

We now consider homogeneous flows. Let us go back to the time series
of dissipation along tracer trajectories in HIT shown in
figure~\ref{fig:F100}. Panel ({\it a}) shows a {\it calm trajectory},
i.e.  a time interval during which dissipation undergoes moderate
fluctuations around the mean. On the other hand, panel ({\it b}) shows
a trajectory that experiences strong intermittency, i.e. the
dissipation undergoes sudden bursts during which its value for a short
time exceeds the average dissipation by several standard deviations
\citep{YeungJFM2001}. Such bursts in dissipation are caused by the
trapping of particles in intense but short-lived vortex structures
\citep{Biferale_trapping2005}, which create very high velocity
gradients and, as shown below, have a distinct influence on the
breakup of strong aggregates.

Panel ({\it c}) in figure \ref{fig:F100} reports the behaviour of the
dissipation along a tracer trajectory in STF. The panel shows the
dissipation over a time interval of $2000\times t_\eta$, from which it can be
 seen that the signal is controlled by two time scales, namely
$t_\eta$ that controls the fast fluctuations and $t_L$ that controls
the slow fluctuations. Magnifying the time series, as done in the
inset of panel ({\it c}), highlights the correspondence of the fast
fluctuations in the synthetic flow to the fine.scale fluctuations in
the homogeneous and isotropic turbulent flow. Also, the time series of
dissipation in the synthetic flow describes a much more regular signal
than that in turbulence, i.e. intermittent bursts and strong
deviations from the mean are absent in this flow. As shown in the next
section, this limits the capability of the STF to break strong
aggregates.

\begin{figure}\begin{center}
\includegraphics[width=8cm]{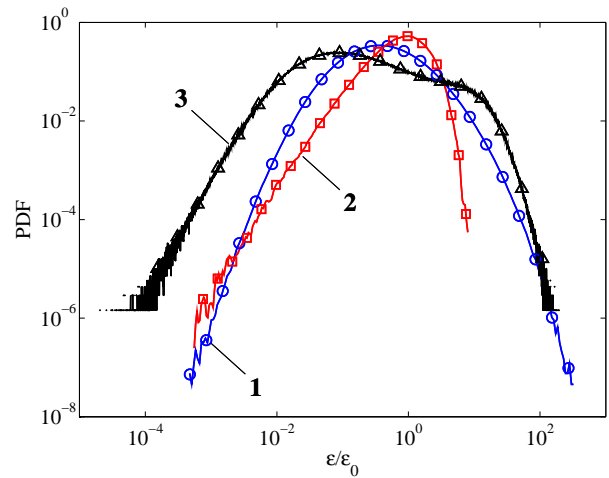}
\caption{\label{fig:F500} Log-log plot of the p.d.f.s of energy
  dissipation in (1) homogeneous and isotropic turbulence, (2)
  synthetic turbulent flow, and (3) channel flow. The vertical axis is
  normalized by $\eps_0$ given in table \ref{tab:characteristics},
  which for the shown flows refers to the volume average.}
\end{center}\end{figure}

Finally, the Eulerian p.d.f.s of the energy dissipation of the homogenous
flows are plotted in figure~\ref{fig:F500}, together with the
dissipation p.d.f. of the CF. In agreement with other studies
\citep{Vedula2001,Yeung_PoF2006}, the dissipation p.d.f. of HIT for
the given Reynolds number exhibits a left tail that is close to
log-normal and a peak value that is slightly smaller than the mean
dissipation. In comparison, the dissipation p.d.f. of the STF is much
narrower. Lastly, the dissipation p.d.f. of the CF is very wide as a
consequence of the non-homogeneity of the flow. The p.d.f. in fact 
exhibits two pronounced shoulders corresponding to the values of
$\eps$ in the bulk (left shoulder) and in the wall regions (right
shoulder).

%
%
\subsection{Breakup rate measurements}
\begin{figure*}\begin{center}
\includegraphics[width=12cm]{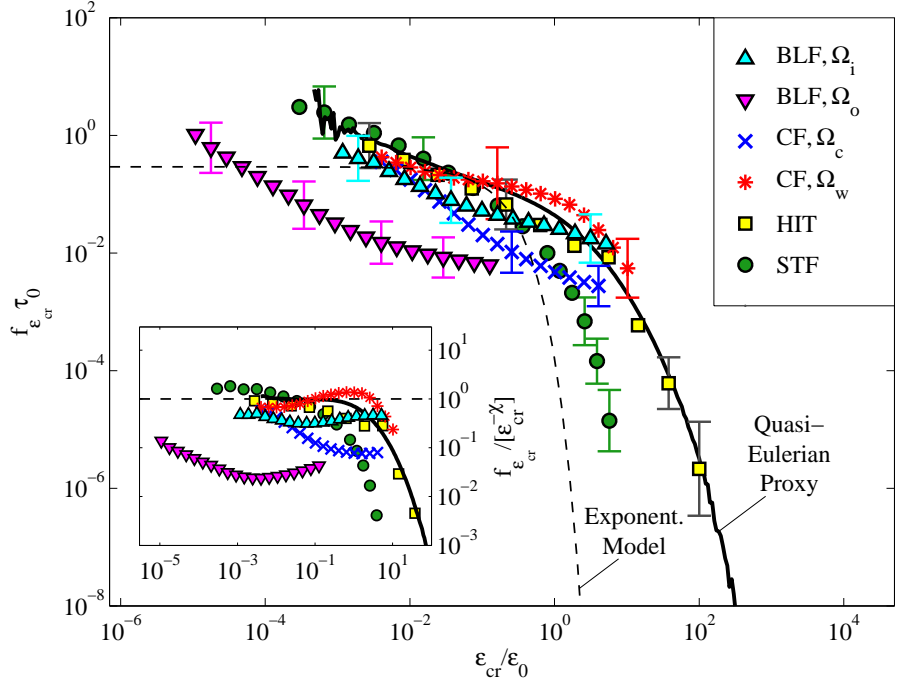}\\
\caption{\label{fig:all_data} Breakup rate as a function of the
  critical dissipation. Symbols: exit-time measurements in a given
  flow configuration and release region (see table
  \ref{tab:characteristics}); the last three points on the right for
  the homogeneous and isotropic turbulence data represent estimates
  from the decay of the number of aggregates according to
  (\ref{eq:NdN}).  Solid line: quasi-Eulerian proxy to homogeneous and
  isotropic turbulence (\ref{eq:Loginov}); dashed line: exponential
  model (\ref{eq:Kusters}).  Inset: Compensated breakup rate
  $f_{\epsCr}/[ \epsCr^{-\chi}]$ with $\chi=0.42$
  \citep{BablerPRE2012}. Symbols have the same meaning as in the main
  axis.}
\end{center}\end{figure*}

We now have all the ingredients needed to measure and rationalize the breakup
rates in turbulent flow upon changing the turbulent intensity and the
mean flow configuration. The results are summarized in figure
\ref{fig:all_data}, which is the major result of this work. In
figure \ref{fig:all_data} we report the breakup rates measured when one changes the flow configuration, the release region and the method
used to estimate $f_{\epsCr}$. We stress that, to the best of our
knowledge, this is the first attempt to perform such a comprehensive
examination of a wide range of flow configurations.

Additionally, we report results from independent predictions, namely the
estimate obtained from quasi-Eulerian measurements given, in
(\ref{eq:Loginov}), and the approximation based on an exponential fit
given in (\ref{eq:NdN}). As expected, the breakup rate generally
decreases with increasing aggregate strength, confirming earlier
results suggesting that large aggregates break faster than small
ones. Remarkably, except for data from the BLF where aggregates were
released outside the boundary layer region (BLF-$\Omega_o$),
the breakup rates of the different flows are quite close
  to each other for small threshold values. We stress that this is
not due to a rescaling of the axis but, rather, a consequence of using
the characteristic dissipation $\eps_0$ and its corresponding
time scale for normalizing the axis. The smaller breakup rates
  for the BLF-$\Omega_o$ case reflect the time it takes for the
  aggregates to be entrained into the boundary layer.

Furthermore, for small $\epsCr$, the breakup rate shows a power-law-like behaviour that is similar for  the different datasets.
Power-law breakup rates have been proposed for describing the
  evolution of the aggregate size distribution in the framework of
  population balance models, where they lead to adequate agreement
  with experiments, at least within a certain range of experimental
  parameters \citep{BablerJCIS2007}. To explore this further, in the
inset of figure \ref{fig:all_data} we show the compensated breakup
rate $f_{\epsCr}/[\epsCr^{-\chi}]$ using a scaling exponent
$\chi=0.42$. The latter corresponds to a fit of the right tail of the
quasi-Eulerian proxy (\ref{eq:Loginov}), which well describes breakup
in HIT, as shown by the solid curve in figure \ref{fig:all_data}.
A distinct plateau can be observed for aggregates released
  close to the wall and for aggregates in homogeneous flows; the
  apparently faster approach to the plateau seen for the wall-bounded
  flows could be due to a more regular dissipation signal at small
  $\eps$ for wall-bounded flows as compared to HIT.  Deviations from
the plateau are only seen for aggregates released in the centre-plane of
the channel (CF-$\Omega_c$) and for aggregates released outside the
boundary layer (BLF-$\Omega_o$), for which the breakup rate for small
$\epsCr$ has a slightly larger scaling exponent. For these release
regions, the aggregates first get entrained by turbulent eddies that
transport them to the wall. During this entrainment the aggregates
gradually experience stronger stress (cf. trajectory A in figure
\ref{fig:F400}). Weak aggregates will therefore, on average, suffer
breakup earlier than stronger ones, which causes the breakup rate for
these release regions to decrease faster with increasing $\epsCr$.

For larger threshold values, a levelling-off in the decrease of the
breakup rate is observed for the wall-bounded flows, i.e. $f_{\epsCr}$
is found to bend upwards as seen in, for instance, the
  BLF-$\Omega_o$ case. This is in contrast to the homogenous flows,
for which $f_{\epsCr}$ shows a strong dropoff at large $\epsCr$
\citep{BablerPRE2012}. The higher breakup rates for wall-bounded flows
are due to the high mean shear close to the wall, which causes
aggregates coming close to the wall to rapidly break up. In the
homogenous flows, strong aggregates are only broken by the rare
excursions of dissipation from the mean caused by intermittency. As
these events are rare, the breakup rate exhibits a superexponential
dropoff for large dissipation. In the STF, where strong and
intermittent excursions from the mean are absent, the dropoff in the
breakup rate occurs at much smaller threshold values than in the case
of three-dimensional turbulence.

The differences between the STF and real HIT for high threshold
values reflects the intriguing dynamics of turbulent fluctuations and the
difficulty of modelling them. Indeed, only for these two cases are the
statistics of aggregate breakup high enough to allow us to assess
the superexponential dropoff and thus to reveal the importance of
turbulence: extremely robust aggregates break only due to the
occurrence of corresponding extremely intense fluctuations, typical of
the intermittent nature of small-scale turbulent flows. Any stochastic
surrogate that does not possess these critical features would severly
underpredict the breakup rate, as is the case for the STF
analysed here.

For very large threshold values, a dropoff in the breakup rate is
also seen for the channel flow where the aggregates are released close
the wall (CF-$\Omega_w$). It represents the situation where the aggregates
are too strong to be broken by the mean shear, and only intense but
rare turbulent fluctuations within the near-wall region are able to
overcome the aggregate strength. A similar dropoff is likely to occur
also for the other cases if trajectories are followed for long enough:
recall that the breakup rate represents the inverse of the
  mean time for which an aggregate survives in the flow. Measuring the small
  breakup rates expected for large threshold values therefore requires
  very long trajectories. Exploring this region of high threshold
values presents a problem for future work.

In addition, figure \ref{fig:all_data} shows the breakup rate for the
exponential model of \citet{Kusters_thesis} (dashed curve).  This
model is based on the simple dimensional assumption that energy
dissipation fluctuations, which govern breakup, have a Gaussian
distribution. As a consequence, the following prediction of the
breakup rate is obtained:
\begin{equation}\label{eq:Kusters}
f^{(K)}_{\epsCr}=\frac{(4/15\pi)^{1/2}}{ (\nu/\langle\eps\rangle)^{1/2} } \exp(-15/2\ \epsCr/\langle\eps\rangle).
\end{equation}
The exponential model predicts a very sharp dropoff at intermediate
threshold values and a constant breakup rate for small threshold
values, which strongly disagrees with the breakup rate found in the
simulations. The discrepancy originates mainly from the simplified
assumption of a Gaussian-like dissipation.

\begin{figure}\begin{center}
\includegraphics[width=8cm]{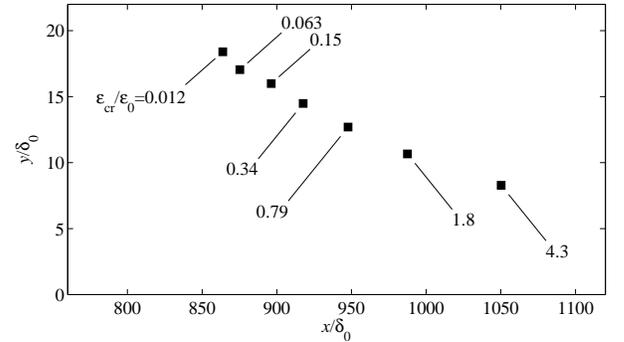}
\caption{\label{fig:F700} Average $(x,y)$-coordinates of the breakup
  position for different threshold values of aggregates released
  in $\Omega_i$ in the boundary layer flow (see figure \ref{fig:bl}).}
\end{center}\end{figure}

\begin{figure}\begin{center}
\includegraphics[width=8cm]{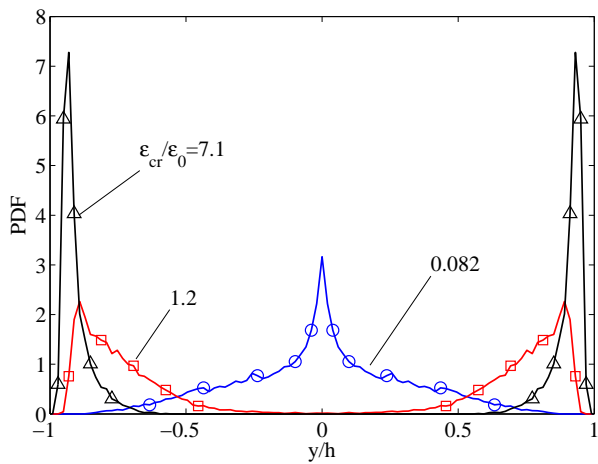}
\caption{\label{fig:F800} Distribution of the wall normal distance
  where breakup occurs in the channel flow for aggregates released in
  $\Omega_c$ (see figure \ref{fig:sketch_CF}). Different curves refer
  to different values of the critical dissipation.}
\end{center}\end{figure}

The observation made from figure \ref{fig:all_data} suggests that weak
aggregates in the wall-bounded flows are broken by turbulent
fluctuations shortly after their release, while on the other hand 
strong aggregates survive for a longer time, during which they move
further downstream where they eventually suffer breakup due to the
mean shear. To explore this further, we examined the spatial location
at which breakup occurs in the wall-bounded flows. Two cases are
considered: aggregates released inside the boundary layer of the BLF
(figure \ref{fig:F700}) and aggregates released in the centre-plane of
the CF (figure \ref{fig:F800}). Figure \ref{fig:F700} shows the
average streamwise and wall-normal coordinates at which breakup occurs
for different threshold values. As can be seen, with increasing
threshold  values the aggregates on average move further downstream and come
closer to the wall before suffering breakup. The average breakup
location for weak aggregates is therefore close to the average location
of where the aggregates were released. Figure \ref{fig:F800} shows the
p.d.f. of the breakup location in the CF for three different threshold
values. It can be seen that weak aggregates predominantly break in the
bulk of the channel close to the point of release, whereas strong
aggregates move further downstream and predominantly break close to
the wall. This observation is important for applications, and might
open a way to tailor turbulent filters with different selection
properties depending on the spatially evolving intensity of the
turbulent background.

%
%
\subsection{Evolution of the number of aggregates}
Strong aggregates can move away from the point of release towards the
high-shear zones close to the walls: this fact has a clear influence
on the breakup behaviour, and leads to the high breakup rates at large
threshold values in wall-bounded flows (cf. figure
\ref{fig:all_data}). This preferential breakup in specific regions of
the flow is also reflected in the time evolution of the number of
aggregates, $N_{\epsCr}(t)$, present in the suspension. From
(\ref{eq:NdN_exit}), it is understood that $N_{\epsCr}(t)$ is
proportional to the cumulative exit-time distribution. As in the
previous section, we limit the discussion to three cases, namely
aggregates released inside the boundary layer in the BLF, aggregates
released in the centre-plane in the CF and aggregates released in
HIT. Not shown is the time evolution in the STF for which the time
evolution of the number of aggregates has the expected result of a
Poisson process.

\begin{figure}\begin{center}
\includegraphics[width=8cm]{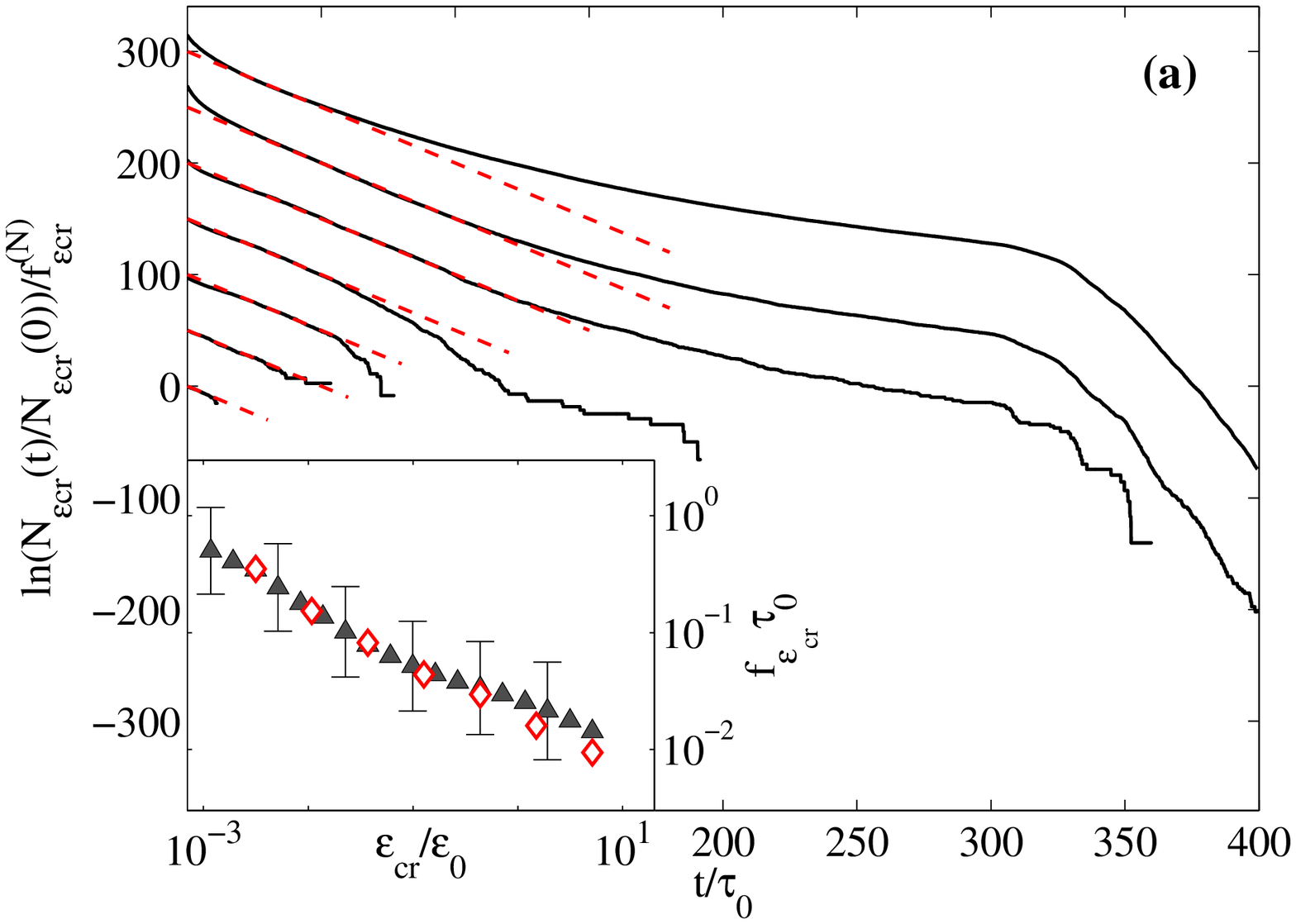} \\
\includegraphics[width=8cm]{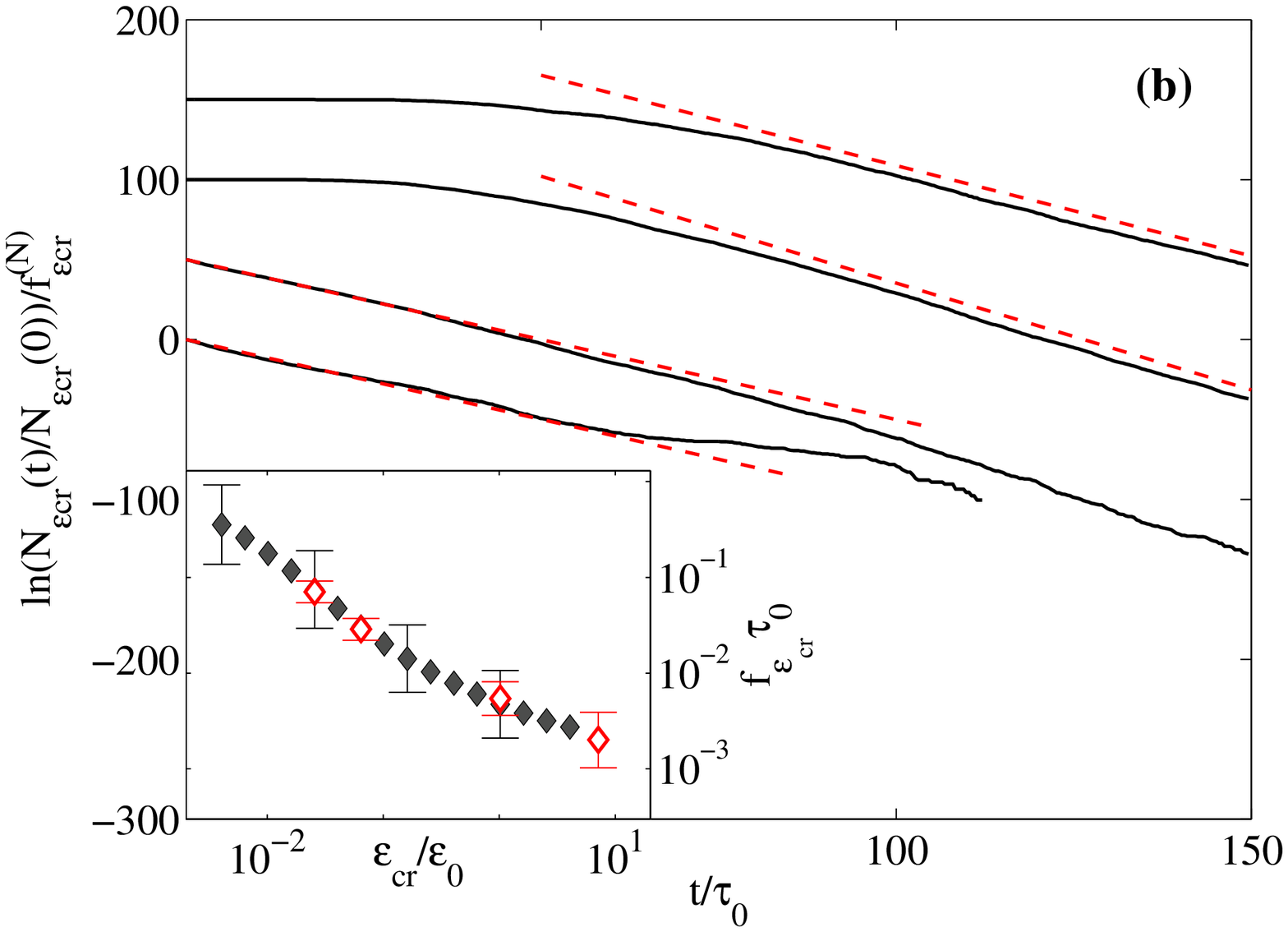} \\
\includegraphics[width=8cm]{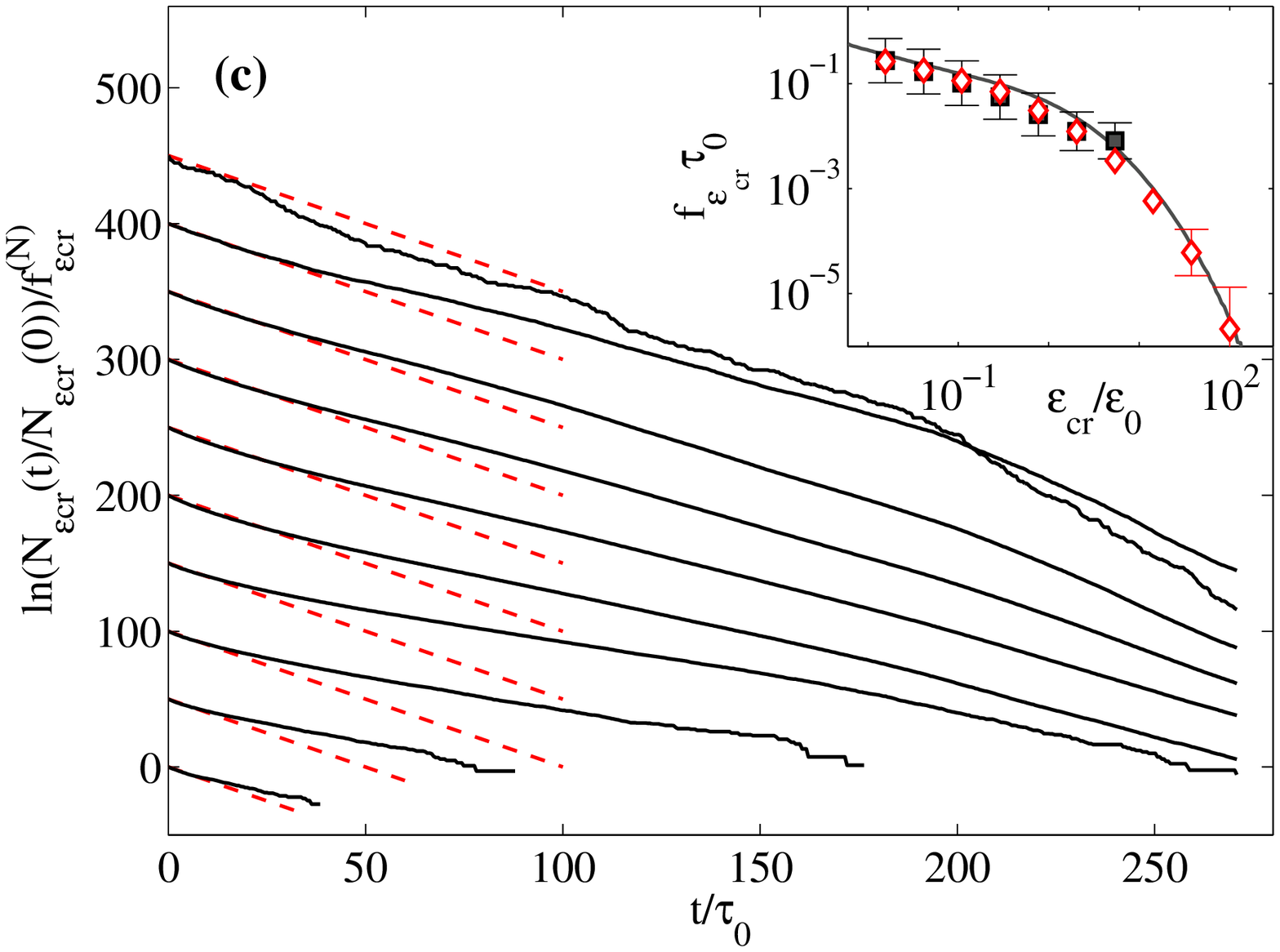}
\caption{\label{fig:F901} Evolution of the number of aggregates in
  semi-logarithmic coordinates for ({\it a}) boundary layer flow with
  aggregates released in $\Omega_i$, ({\it b}) channel flow with
  aggregates released in $\Omega_c$, and ({\it c}) homogeneous and
  isotropic turbulence.  Main axes: $\ln(N_{\epsCr}(t)/N_{\epsCr}(0))$
  for different threshold values normalized by the slope
  $f^{(N)}_{\epsCr}$ of the pure exponential decay. For clarity, the
  curves are shifted upwards by a fixed increment such that $\epsCr$
  increases from bottom to top. The dashed lines indicate the linear
  regions used to fit $f^{(N)}_{\epsCr}$.  Inset: Breakup rate as a
  function of critical dissipation obtained from exit-time
  measurements (solid symbols) and linear fits to $\ln N_{\epsCr}(t)$
  (open symbols). The former is the same as data plotted in figure
  \ref{fig:all_data}. The solid line in inset ({\it c}) shows the
  quasi-Eulerian proxy (\ref{eq:Loginov}).  }
\end{center}\end{figure}

In figure~\ref{fig:F901}({\it a}), the evolution of the number of
aggregates released inside the boundary layer in the BLF is
displayed. The figure shows $N_{\epsCr}(t)$ in semilogarithmic
coordinates, with the different curves corresponding to different
threshold values. It is clear that for small threshold values (lower
curves in figure \ref{fig:F901}({\it a}), the number of aggregates
decays exponentially as $N_{\epsCr}(t) \simeq N_0
\exp(-f^{(N)}_{\epsCr}\,t)$. Deviations from the exponential decay
observed at later times are due to statistical noise as the number of
aggregates is already very small. The slope $f^{(N)}_{\epsCr}$, as
suggested by (\ref{eq:NdN}), provides an estimate of the breakup rate.
The exponential decay represents the case where the aggregates are
broken by uncorrelated turbulent fluctuations in the vicinity of the
point of release. 

On the other hand, for large threshold values (upper
curves in figure \ref{fig:F901}({\it a}), the evolution of the number
of aggregates shows a different pattern: after an exponential decay at
earlier times, a relaxation sets in at intermediate times which
eventually turns into an abrupt decrease at later times. Of these
three stages, the relaxation following the exponential decay of
$N_{\epsCr}(t)$ is caused by aggregates surviving early breakup and
moving away from the point of release. However, later, when these
aggregates come close to the wall, they suffer abrupt breakup as
represented by the third stage.

The good news here is that despite such non trivial time evolution,
estimating the breakup rate from the linear segments of $\ln
N_{\epsCr}(t)$ provides a reasonable approximation. This is shown in
the inset of figure \ref{fig:F901}({\it a}) where we compare the
breakup rate measured by the mean exit time, as plotted in figure
\ref{fig:all_data}, with the estimation from the linear segments. The
latter is very close to the former, which implies that for the threshold 
values considered, the rate of breakup in the boundary layer
flow is controlled by the early breakup events in the vicinity of the
point of release of the aggregates.\\

The evolution of the number of aggregates released in the centre-plane
of the CF is shown in figure \ref{fig:F901}({\it b}). As in the BLF, 
for small threshold values (lower curves in figure
\ref{fig:F901}({\it b})), the number of aggregates decays
exponentially, implying that the aggregates are broken by short-time
correlated turbulent fluctuations in the vicinity of the point of
release. On the other hand, for large threshold values (upper curves
in figure \ref{fig:F901}({\it b})), the evolution of $N_{\epsCr}(t)$
is delayed and its decay sets in only after the aggregates have been
in the flow for a certain time. This delay reflects the time it takes
for the aggregates to get entrained into turbulent eddies which
transport them to the higher-shear regions close to the wall where
they eventually suffer breakup. The breakup rate estimated from
fitting the linear segments of $\ln N_{\epsCr}(t)$ is shown in the
inset of figure \ref{fig:F901}({\it b}), together with the exact
breakup rate obtained from exit-time measurements as plotted in figure
\ref{fig:all_data}.  Good agreement with the exact breakup rate is
observed also in this case.

Lastly, the evolution of the number of aggregates in HIT is shown in
figure \ref{fig:F901}({\it c}). In contrast to the wall-bounded flows,
no qualitative difference in the decay of $N_{\epsCr}(t)$ for
different threshold values is seen, and for all threshold values an
exponential decay is observed, as indicated by the dashed lines. The
breakup rate estimated from this initial decay is shown in the inset
of figure \ref{fig:F901}({\it c}), together with the exact breakup
rate from figure \ref{fig:all_data}. The latter is available up to
threshold values $\epsCr/\langle\eps\rangle\sim 5$; beyond this value,
exit times are large compared with the duration of our numerical
simulation which precludes exact measurements of the breakup rate. As
can be seen from the inset, for threshold values smaller than
$\epsCr/\langle\eps\rangle\sim 5$ the approximated breakup rate is
very close to the exact breakup rate, while for larger threshold
values it is in good agreement with the {\it quasi}-Eulerian proxy
shown by the solid curve. This close agreement indicates that breakup
in HIT resembles breakup in short-time correlated force fields, which
justifies modelling breakup as a first-order rate process.

%
%
\section{Conclusions}

We have reported the first systematic study concerning the estimation of
the breakup rate of small aggregates in fully developed turbulence upon
changing both the flow configuration (bounded and unbounded), and the
injection region (relevant only for the bounded-flow cases). Also, we have
discussed theoretical and phenomenological ideas concerning the
definition of the breakup rate in terms of the so-called exit times
measured along the trajectories of all aggregates, or in terms of
other proxies, such as breakup rates defined by means of fully Eulerian
quantities or by using a fast decorrelation hypothesis along Lagrangian
trajectories. Our main approximations are the assumptions that breakup
occurs instantaneously once the dissipation at the position of the
aggregate exceeds a predefined threshold value and that aggregates
behave like tracers with only a one-way coupling with the flow (i.e. no
inertia and no feedback on the flow). In future work, the former
restriction could be overcome by considering certain time-relaxation
properties of the aggregate backbone; the latter can be relaxed by
considering inertial aggregates (still one-way coupling).

We have found that breakup is typically the result of two competing
  effects: a systematic influence of the mean turbulent profile,
overlaid by intermittent and bursty events induced by turbulent
fluctuations. In turbulent regions dominated by small dissipation
events, important for large and easy-to-break aggregates, the breakup
rate shows a similar pattern in all flows considered. In particular, we
found that the breakup rate in the different flows exhibits a
qualitatively similar power-law scaling. This can be explained by noticing
that weak aggregates are broken by turbulent fluctuations in the
vicinity of the point of release. As the local properties of
turbulence at the injection point are expected to be similar, in
dimensionless units, the breakup rate assumes similar values.

On the other hand, breakup rates driven by large dissipation events
are significantly different between the four flows. Compared to
homogeneous isotropic turbulence, the bounded flows lead to
persistently high breakup rates even for large values of the
threshold dissipation. This is due to the fact that in non-homogeneous
flows, aggregates can be broken also by the mean flow if they travel
enough to reach regions close to the boundary. On the contrary, the
synthetic turbulent flow shows very small breakup rates for large
threshold dissipations, due to the absence of both a mean profile and
intense intermittent fluctuations characteristic of realistic
homogeneous and isotropic turbulent flows.

The study presented here can be viewed as a first step towards the
systematic development of models for aggregation kernels and breakup
rates, to be used in spatially distributed population balance and
compartment models. 
Furthermore, it helps us devise experiments for measuring breakup rates.
Experimental approaches to measuring breakup rates can be divided into two types:
One type of approach measures the time evolution of the aggregate size distribution from which the breakup rate is deduced, for example by means of a population balance equation (PBE) model. The difficulty of this approach is that the time resolution for measuring the size distribution of such small aggregates is of the order of seconds at best. 
Also, as the resolution of the measured size distributions is relatively coarse, the PBE model typically needs a predescribed function for the breakup rate. Despite these difficulties, results from this approach have yielded valuable insights into aggregate breakup: it was found that under certain conditions, a power-law breakup rate is in agreement with experimental data \citep{BablerJCIS2007}. A power-law breakup rate in the limit of small threshold dissipation (corresponding to large aggregates) was also found in our study. 
In the second type of approach, discrete breakup events of aggregates are observed using, for example, stereoscopic microscopy together with particle velocity tracking. Such an approach has been pursued by L{\"u}thi and coworkers \citep{Saha_thesis}. However, the experiments turned out to be fairly tedious in terms of both following the aggregate and observing its breakup. Hence, the methodology of and insight provided by our work can serve as valuable input for the experimentalist. 
\\

Computer time provided by SNIC (Swedish National Infrastructure for
Computing) is gratefully
acknowledged. DNS of homogenous and isotropic turbulence were performed at HPC Centre CINECA (Italy).  M.U.B. was financially supported by the Swedish Research
Council VR (grant no. 2012-6216). L. Biferale acknowledges partial funding
from the European Research Council under the European Community's
Seventh Framework Program, ERC Grant Agreement no. 339032. EU-COST
action MP0806 is kindly acknowledged.


\bibliography{Referenser}

\end{document}